\documentclass[reprint,prl,nofootinbib,superscriptaddress]{revtex4-2}
\usepackage[T1]{fontenc}
\usepackage[utf8]{inputenc}
\usepackage{lmodern}
\usepackage{amsmath,amssymb}
\usepackage{graphicx}
\graphicspath{{figs/}}

\begin{document}
\raggedbottom

\title{Metric random tilings: counted geometric disorder and its diffraction}

\author{Xinyan Wu}
\author{Qingqing Ji}
\email{jiqq@shanghaitech.edu.cn}
\affiliation{State Key Laboratory of Quantum Functional Materials, School of Physical Science and Technology, ShanghaiTech University, Shanghai 201210, People's Republic of China}

\begin{abstract}
Random tilings gain entropy through the many ways fixed shapes can meet. What disorder remains possible when every face keeps its neighbors and every edge its direction? We introduce metric random tilings, a new class of disordered statistical models in which shared edges take discrete lengths that must close every face. For three equally spaced lengths, four of the eleven uniform tilings admit a number of geometries that grows exponentially with area. On the honeycomb this multiplicity survives even when the boundary and inventory of each flower patch are prescribed. Breaking the equal spacing lowers the estimated honeycomb entropy from about half a Boltzmann constant per face to about a fifth. Shifting all lengths together preserves every pattern and its probability, yet changes the coherent diffraction intensity by nearly three orders of magnitude on matched honeycomb samples. Metric random tilings thus realize bulk disorder at fixed topology and reveal how strongly its diffraction signature can vary at fixed configurational entropy.
\end{abstract}

\maketitle

\section{Introduction}\label{sec:intro}

Suppose every pair of neighbors and the direction of every bond in a two-dimensional solid are known. How much configurational entropy could remain? An isolated polygon can change shape at fixed edge directions. In a tiling, however, each interior edge belongs to two faces, so changing its length to close one constrains the next. Bulk entropy requires choices that survive this coupling at finite density.

\begin{figure*}
\includegraphics[width=0.97\textwidth]{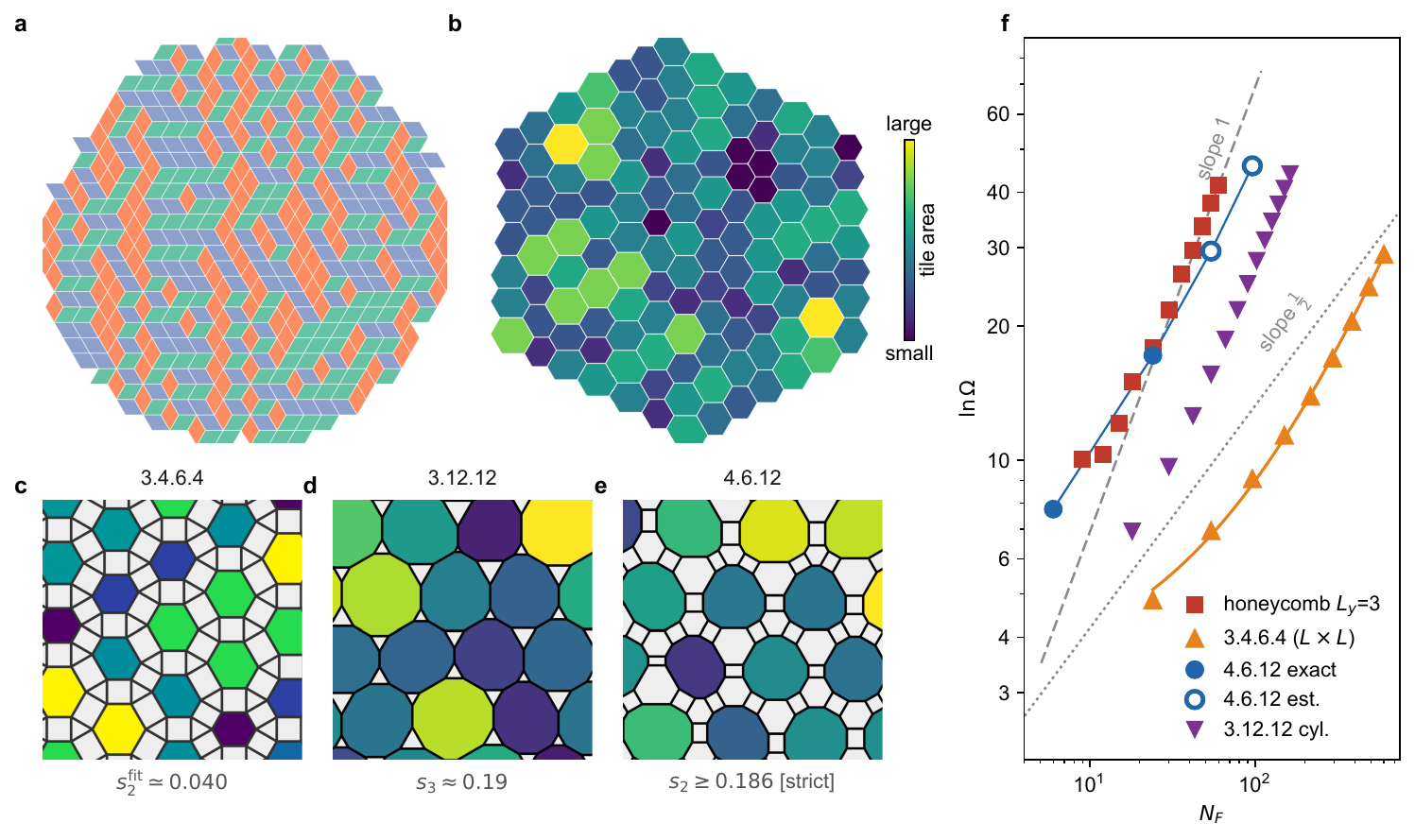}
\caption{Sources of tiling entropy. \textbf{a}, Fixed lozenges with varying contacts. \textbf{b--e}, Honeycomb, 3.4.6.4, 3.12.12 and 4.6.12 configurations colored by face area. Labels give fitted bulk estimates (binary \textbf{c}, ternary \textbf{d}) and a width-three strip bound (\textbf{e}). \textbf{f}, Exact counts (filled) and calibrated estimates (open), with area and perimeter guides.}
\label{fig1}
\end{figure*}

Much of tiling theory starts from the complementary problem of arranging prescribed shapes. In Kepler's uniform tilings, regular polygons meet in the same arrangement around every vertex~\cite{Kepler1619,GrunbaumShephard}. Matching rules for fixed shapes can enforce aperiodic order~\cite{Penrose1979,DeBruijn1981}, and sharp diffraction can survive the loss of periodicity~\cite{Mackay1982,Shechtman1984,Levine1984}. When many arrangements are allowed, tile contacts themselves can carry entropy (Fig.~\ref{fig1}a)~\cite{Elser1985,Henley1991,OxborrowHenley1993}, as in square--triangle and octagonal random tilings~\cite{Widom1993,deGierNienhuis1996}. In the honeycomb of Fig.~\ref{fig1}b we reverse what is prescribed, keeping the neighbors of each face fixed while allowing its shape to vary.

We call these ensembles \emph{metric random tilings}. Here connectivity and edge directions come from a periodic reference, while compatible discrete lengths determine the positions of the sites themselves. The eleven uniform tilings let us test when this freedom yields bulk entropy (SI S1, Table S1).

Metric random tilings seek configurational entropy in compatible edge lengths, drawing on the principle that locally constrained variables can sustain bulk entropy~\cite{Pauling1935,Lieb1967,Kasteleyn1961}. Unlike Wang-tile colors~\cite{Wang1961,Chen2016}, the matching labels are Euclidean distances shared by adjacent faces. The geometric constraints connect the model to rigidity theory~\cite{Calladine1978,GuestFowler2014} and edge-length descriptions of discrete geometry~\cite{Regge1961,Bittner1999}. The statistical question is how many assignments from a finite alphabet close all faces simultaneously.

\section{Counting geometries at fixed connectivity}\label{sec:count}

Assign each shared edge a length $\ell_e$ from a finite set $A\subset\mathbb R_{>0}$. Following the boundary of face $f$ back to its starting vertex requires
\begin{equation}
\sum_{e\in\partial f}\ell_e\,\hat{\mathbf t}_{f,e}=\mathbf 0,
\end{equation}
where $\hat{\mathbf t}_{f,e}$ is the unit tangent along the oriented boundary. Collecting the face equations on a domain $D$ with $F_D$ faces and edge set $E_D$ gives the closure matrix $C_D$, the ensemble and its count,
\begin{equation}
\mathcal C_D(A)=\{\boldsymbol\ell\in A^{E_D}:C_D\boldsymbol\ell=0\},
\qquad
\Omega_D(A)=|\mathcal C_D(A)|.
\end{equation}

At equal weights the entropy per face, in units of $k_{\mathrm B}$, is $\ln\Omega_D/F_D$. Consider a honeycomb of edge length 2. Shorten the six edges of one hexagon to 1 and lengthen the six radiating from its corners to 3. On each neighboring face the two outer tangents of the affected triplet sum to the middle tangent, so closure is preserved. All 18 outer edges of this seven-face patch retain length 2. Interchanging short and long edges gives a second interior with the same boundary and inventory.

Placing patches three cells apart on an $L\times L$ torus gives $\lfloor L/3\rfloor^2$ independent binary choices and hence at least $(\ln 2)/9$ per face in the large-system limit (SI S2, Fig.~S1a). Each flower thus stores an independent bit at fixed boundary and composition. This local rearrangement, geometrically dual to the triangular-lattice wheel move in granular force network ensembles~\cite{Tighe2005,CaoMilner2014}, provides the building block for extensive metric entropy.

Independent patches also exist on 3.4.6.4, 3.12.12 and 4.6.12 (Fig.~\ref{fig1}c--e). On 3.4.6.4 patches can even share a face: their changed edges are disjoint and each set of changes preserves closure. The ternary lower bounds are $(\ln 2)/24$, $(\ln 2)/18$ and $(\ln 2)/36$ per face, respectively (SI S2, Fig.~S1b--d).

Figure~\ref{fig1}f tracks $\ln\Omega$ against face count on strips and tori. For 3.4.6.4, the area coefficient fitted to exact binary torus counts gives $s_2\approx0.040$ per face (Fig.~\ref{fig1}c). For 3.12.12, extrapolating ternary transfer-matrix rates across widths gives $s_3\approx0.19$ (Fig.~\ref{fig1}d). For 4.6.12, calibrated Knuth sampling extends the ternary torus series. Fig.~\ref{fig1}e gives a binary lower bound of 0.186 per face at strip width three (SI S3, Figs.~S2 and S3). Together, the constructions and counts identify four of Kepler's eleven uniform tilings that support extensive metric entropy at three equally spaced lengths. What does face closure allow on the remaining seven?

\section{How face closure limits metric entropy}\label{sec:flex}

\begin{figure*}
\includegraphics[width=0.72\textwidth]{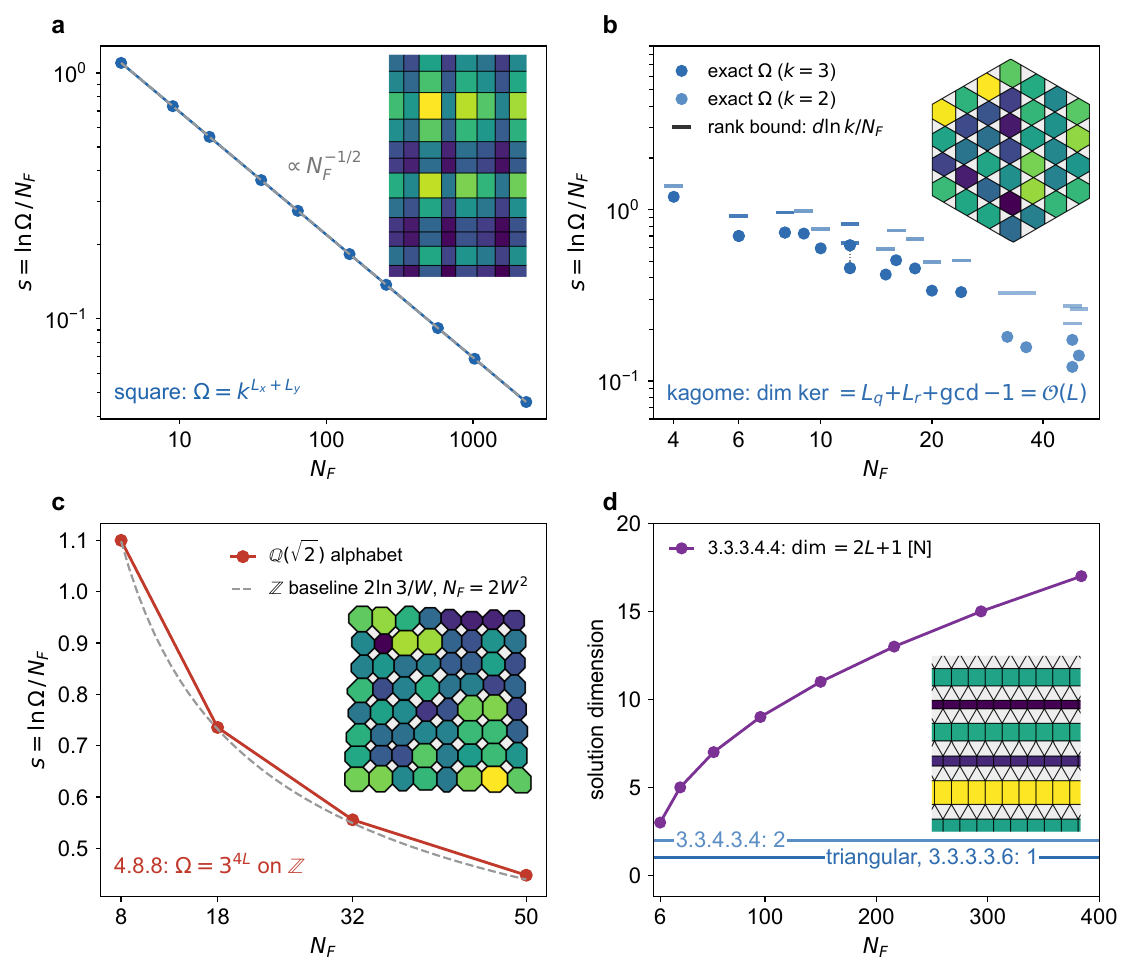}
\caption{Geometric and arithmetic restrictions on bulk metric entropy. \textbf{a}, Opposite-edge equality propagates through square-lattice rows and columns. \textbf{b}, Kagome counts and the rank bound from three lines of surviving modes. \textbf{c}, Integer lengths on 4.8.8 obey a perimeter law despite an extensive real kernel. Points show a finite alphabet in $\mathbb Q(\sqrt{2})$ (W=2--5). \textbf{d}, Real solution dimensions on finite tori: one global scale for the triangular and snub hexagonal tilings, two lengths for the snub square, and the $2L+1$ sequence for 3.3.3.4.4 (rank checks in SI S4.1). Insets show face areas.}
\label{fig2}
\end{figure*}

First let the edge lengths vary continuously. Let $d$ count the independent lengths left by closure; those $d$ edges determine all the others. Restricting each to $k=|A|$ values therefore gives $\Omega\le k^d$: bulk entropy requires $d$ to grow with area.

On the square lattice, opposite-edge equality propagates each length along a row or column, giving $\Omega=k^{L_x+L_y}$ (Fig.~\ref{fig2}a; SI S4.1). On the kagome lattice each triangle has a single length. Fourier transforming these triangle lengths over the periodic cells shows that closure permits modes on only three lines of wavevectors. Thus $d$ grows only with linear size (Fig.~\ref{fig2}b; SI S4.2). Closure leaves just one overall scale on the triangular and snub hexagonal tilings, and two independent lengths on the snub square. For 3.3.3.4.4 the computed $d$ grows linearly with system size (Fig.~\ref{fig2}d; SI Table S2).

On the truncated-square tiling (4.8.8), an $L\times L$ torus admits $2L^2+2$ independent real length variables. For integer lengths one must count the integer points in this solution space~\cite{Beck2004,McAllisterWoods2005}. Axial and diagonal edges contribute rational and $\sqrt{2}$ terms to octagon closure. With integer lengths these terms must cancel separately, forcing opposite edges to agree along $4L$ independent strands. Hence $\Omega=k^{4L}$ (Fig.~\ref{fig2}c; SI S4.3). A suitable ternary alphabet in $\mathbb{Q}(\sqrt{2})$ yields higher counts at the tested sizes (Fig.~\ref{fig2}c; SI S4.3).

Bulk entropy can therefore be suppressed in two ways. Closure can leave too few independent lengths, or the allowed values can realize too few of the geometries that closure permits. On the honeycomb, where bulk entropy survives, how does its magnitude depend on the spacing between allowed lengths?

\section{Which length patterns close on the honeycomb?}\label{sec:arith}

\begin{figure*}
\includegraphics[width=0.97\textwidth]{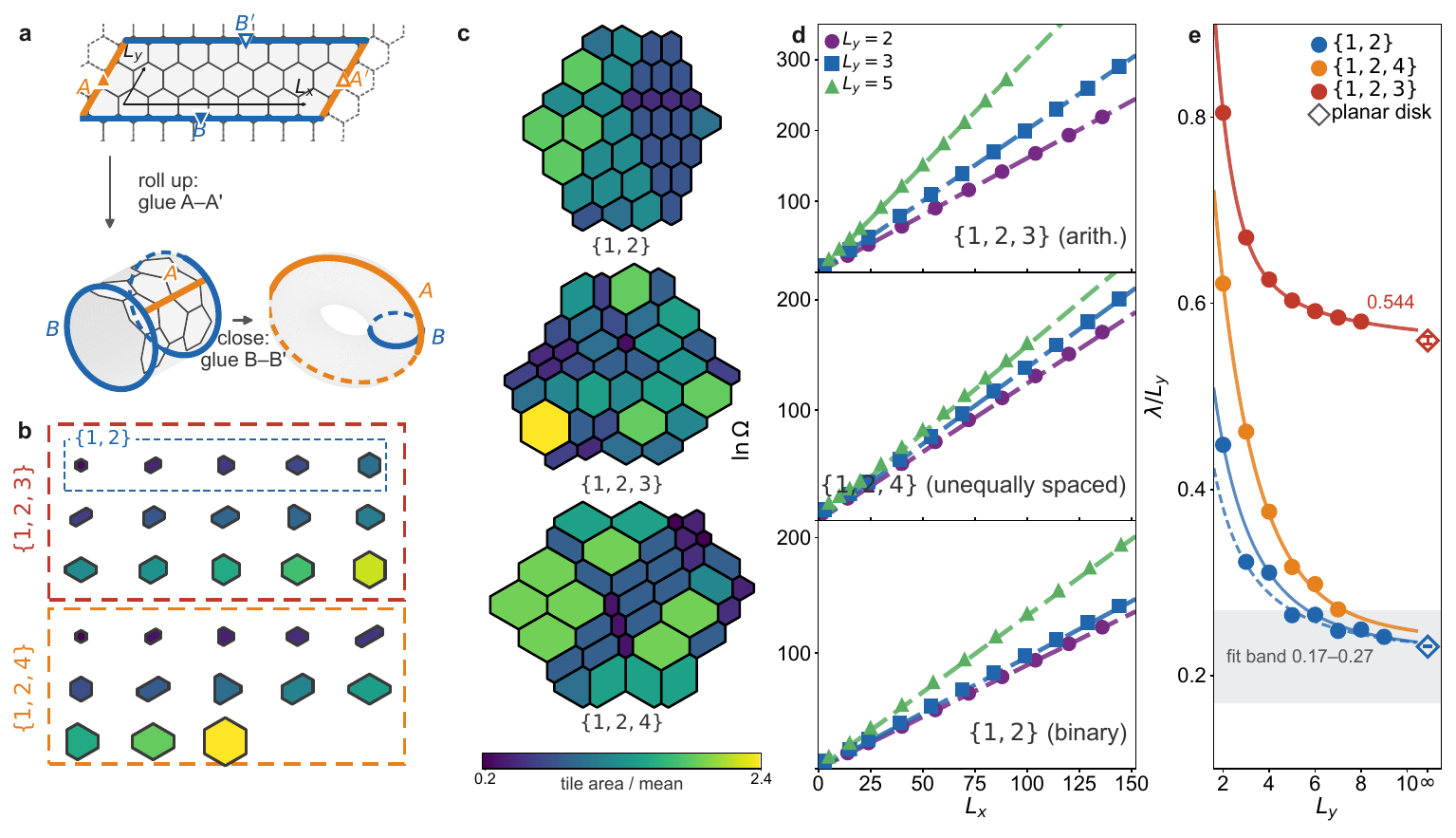}
\caption{Honeycomb length arithmetic. \textbf{a}, Standard embedding; counts use the sheared periods of SI S1. \textbf{b,c}, Closed hexagons up to dihedral symmetry and sample configurations. Boxes indicate binary shapes. \textbf{d}, Periodic counts at selected resonant lengths and fixed-width asymptotes. \textbf{e}, Width-normalized strip rates and extrapolations. Open diamonds are area coefficients fitted to exact disk counts (binary, equally spaced).}
\label{fig3}
\end{figure*}

Gluing opposite boundaries of a honeycomb patch gives a counting torus, illustrated schematically in Fig.~\ref{fig3}a (counting periods in SI S1). Which hexagons can tile it? Closure requires three directed differences between opposite edges to agree. Two lengths, $\{1,2\}$, already admit five hexagon shapes up to rotations and reflections. Adding a third length gives fifteen shapes for $\{1,2,3\}$ but only thirteen for $\{1,2,4\}$ (Fig.~\ref{fig3}b,c). In the equally spaced triple, $2-1=3-2$ lets different edge pairs supply the same difference. In the other two sets each nonzero difference fixes its ordered pair. Changing the ternary spacing eliminates the flower's neighboring-face profiles and cuts the $5\times3$ torus count from 185\,523 to 19\,683.

How do these local compatibility rules govern the global configuration count? Closing a width-$W$ strip after $L$ columns gives $\Omega=\operatorname{Tr}(T_W^L)$. Contributions from eigenvalue families can cancel in this trace at particular periods. For both $\{1,2\}$ and $\{1,2,4\}$, coprime periods give the perimeter count $k^{L_x+L_y+1}$ throughout the tested grid. Yet at $20\times5$ the binary count exceeds that prediction by a factor of $7.7\times10^4$ (Fig.~\ref{fig3}d; SI S5.1, Fig.~S4).

The spectral radius gives the long-strip growth rate $\lambda(W)=\ln\rho(T_W)$. Extrapolating $\lambda(W)/W$ gives about 0.2 nats per face for binary lengths and for $\{1,2,4\}$ (0.17--0.27 across fits). The equally spaced triple has the highest rate at every common width and extrapolates to 0.544 (free fit, range 0.53--0.58). Exact counts on open disks independently give fitted area coefficients of 0.231(1) for binary lengths and 0.560(4) for the equally spaced triple (Fig.~\ref{fig3}e; SI S5.2).

The length set thus enters the statistical mechanics twice, through the number of choices per edge and through the arithmetic that determines their compatibility. Changing the relative gaps can alter those compatibility rules. A common shift preserves them: since the unit-edge reference satisfies $C_D\mathbf 1=0$, adding $c$ to every edge length gives a bijection between admissible configurations, provided all shifted lengths remain positive. The uniform ensembles therefore have identical pattern probabilities and entropy. What, then, changes physically when the metric changes but the entropy does not?

\section{Diffraction at fixed configurational entropy}\label{sec:diff}

\begin{figure*}
\includegraphics[width=0.97\textwidth]{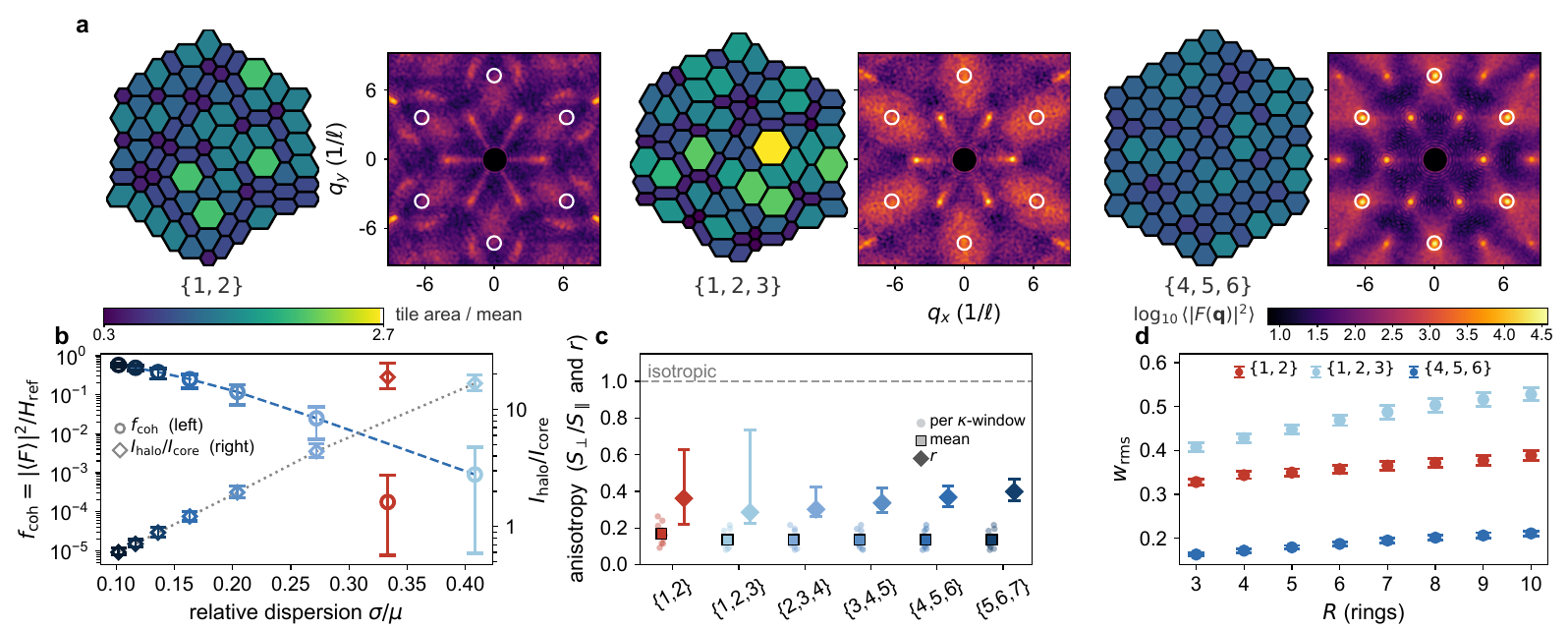}
\caption{Diffraction at fixed entropy. \textbf{a}, Configurations and mean intensity maps for binary lengths and two equally spaced triples. Circles mark the reference second shell. \textbf{b}, Normalized coherent intensity and halo-to-core ratio versus relative length dispersion. Blue points follow twenty matched 169-face configurations through the length sets (left to right) $\{7,8,9\}$, $\{6,7,8\}$, \ldots, $\{1,2,3\}$. Independent binary samples (red) from two parent torus sizes gave coherent fractions of $1.8\times10^{-4}$ and $1.0\times10^{-2}$. Bars show within-run snapshot-bootstrap intervals (SI S6.1). \textbf{c}, Displacement power ratios $S_\perp/S_\parallel$ and halo second-moment ratios $r=\sigma_\perp^2/\sigma_\parallel^2$ for the binary set and five equally spaced triples. \textbf{d}, Displacement roughness versus patch radius.}
\label{fig4}
\end{figure*}

We follow twenty 169-face honeycomb configurations through successive shifts, measuring positions in units of the mean bond length and obtaining their reciprocal-space intensity maps by Fourier transformation. As the length set shifts from $\{1,2,3\}$ to $\{4,5,6\}$, diffraction peaks emerge from the diffuse background (Fig.~\ref{fig4}a; SI S6.1). Each pattern keeps the same weight across the shift. The comparison therefore isolates how the spatial embedding affects diffraction at fixed entropy.

At six reference spots we average the complex amplitude over configurations before squaring. From $\{1,2,3\}$ to $\{7,8,9\}$ the resulting coherent intensity rises from 0.000901 to 0.577 of the ideal reference, a factor of about 640 (Fig.~\ref{fig4}b). The mean total intensity rises 102-fold, while the integrated halo-to-core ratio falls from 16.69 to 0.60. Scattering weight thus concentrates near the peaks, and a larger share of their intensity is coherent.

The sharpening follows from how a length shift moves each vertex. Along a path to vertex $v$, the added edge vectors sum to $c\mathbf R_v$, where $\mathbf R_v$ is its position in the unit-edge reference. After registration and mean-length normalization, the displacement from the reference obeys
\begin{equation}
\mathbf u(\boldsymbol\ell+c\mathbf 1)
=\frac{\bar\ell}{\bar\ell+c}\,\mathbf u(\boldsymbol\ell),
\end{equation}
where $\bar\ell$ is the original mean bond length. For $c>0$ this factor contracts the displacements, narrowing the phase spread. For the two largest shifts ($c=5$ and 6), a Debye--Waller-type estimate~\cite{Debye1913} based on the nominal relative dispersion of the length alphabet reproduces the coherent intensity to within 1\%. At the unshifted endpoint it underestimates the intensity by a factor of 5.8 for the 169-face samples (SI S6.2; size dependence in SI S6.3), suggesting that relative length dispersion alone does not fully capture the phase fluctuations of strongly disordered metric configurations.

Does the recovery of coherent peaks alter the anisotropy of metric disorder? The contraction scales longitudinal and transverse displacement amplitudes equally, preserving their power ratio at each wavevector within a configuration. The ensemble power ratio varies by a factor of 2.7 across the resolved wavevectors but agrees within 0.5\% across five translated sets at fixed wavevector. Longitudinal power dominates, consistent with the radial elongation of the halos (Fig.~\ref{fig4}c). The measured halos nevertheless become less anisotropic as the nearly isotropic tails of the strengthening coherent peaks gain weight (SI S6.2). In real space, the roughness of these displacement fields grows with patch radius, with both logarithmic and shallow power-law growth compatible with the accessible range (Fig.~\ref{fig4}d; SI S6.2--S6.3).

\section{Discussion}\label{sec:disc}

In metric random tilings, length fluctuations displace scattering centers without changing their connectivity, providing a metric counterpart to the phason diffuse scattering of conventional random tilings~\cite{StrandburgTangJaric1989,Socolar1986,NewmanHenley1995,MihalkovicHenley2004}. Such disorder could be realized in frameworks with locked connectivity and directions, using rigid linkers of discrete lengths. Their identification would combine diffuse scattering with local probes of these characteristics, as in disordered crystalline frameworks~\cite{Meekel2023,Meekel2024}. Energetic selection can further reweight these metric configurations~\cite{KeenGoodwin2015,SimonovGoodwin2020}, as illustrated by a field conjugate to total edge length (SI Fig. S5).

Our length-shift construction shows how coherent peaks strengthen without changing pattern probabilities or entropy, complementing entropy variation at fixed diffraction~\cite{BaakeGrimm2009} and extending length-dependent diffraction at fixed probabilities~\cite{BaakeHoffe2000} to a planar system with coupled closure constraints. Metric random tilings thus establish a statistical mechanics of fluctuating shapes at fixed topology. A structure can retain the same connectivity in every microstate, yet store bulk entropy in the very distances that determine its diffraction pattern.

\section*{Data and code availability}

{\raggedright Configuration counts, diffraction outputs, ensembles, and the counting, sampling, validation, and figure-generation scripts are available in the companion repository \mbox{https://github.com/shtusynelab/metric-random-tiling}, archived on Zenodo (DOI: \mbox{10.5281/zenodo.24156133}, concept DOI).\par}

\section*{Author contributions}

X.W. performed the research and developed the counting and analysis pipeline; Q.J. conceived and supervised the research and wrote the manuscript.

\section*{Acknowledgments}

This work was financially supported by the National Key R\&D Program of China (2024YFA1410500), the Science and Technology Commission of Shanghai Municipality (25JC3200500), and the Startup Fund and the Double First-Class Initiative Fund of ShanghaiTech University.

\section*{Competing interests}

The authors declare no competing interests.

\end{document}